\documentclass[twocolumn,superscriptaddress,showpacs,prl]{revtex4}
\usepackage[dvips]{graphicx,color}
\usepackage{delarray}
\usepackage{amsmath}
\usepackage{bm}

\newcommand{\braket}[2]{\langle {#1} | {#2} \rangle}
\newcommand{\ket}[1]{| { #1} \rangle}
\newcommand{\bra}[1]{ \langle {#1}  |}

\begin{document}
\title{Optimal entanglement generation for efficient hybrid quantum repeaters}

\author{Naoya Sota}
\affiliation{Department of Materials Engineering Science, Graduate school of Engineering Science,
Osaka University, Toyonaka, Osaka 560-8531, Japan}
\affiliation{CREST Research Team for Photonic Quantum Information, 4-1-8 Honmachi, Kawaguchi,
Saitama 331-0012, Japan}

\author{Koji Azuma}
\email{azuma@qi.mp.es.osaka-u.ac.jp}
\affiliation{Department of Materials Engineering Science, Graduate school of Engineering Science,
Osaka University, Toyonaka, Osaka 560-8531, Japan}
\affiliation{CREST Research Team for Photonic Quantum Information, 4-1-8 Honmachi, Kawaguchi,
Saitama 331-0012, Japan}

\author{Ryo Namiki}
%\email{}
\affiliation{Department of Physics, Graduate School of Science, 
Kyoto University, Kyoto 606-8502, Japan}

\author{\c{S}ahin Kaya \"Ozdemir}
%\email{}
\affiliation{Department of Materials Engineering Science, Graduate school of Engineering Science,
Osaka University, Toyonaka, Osaka 560-8531, Japan}
\affiliation{CREST Research Team for Photonic Quantum Information, 4-1-8 Honmachi, Kawaguchi,
Saitama 331-0012, Japan}
\affiliation{ERATO Nuclear Spin Electronics Project, Sendai 980-8578, Japan}

\author{Takashi Yamamoto}
%\email{}
\affiliation{Department of Materials Engineering Science, Graduate school of Engineering Science,
Osaka University, Toyonaka, Osaka 560-8531, Japan}
\affiliation{CREST Research Team for Photonic Quantum Information, 4-1-8 Honmachi, Kawaguchi,
Saitama 331-0012, Japan}

\author{Masato Koashi}
%\email{}
\affiliation{Department of Materials Engineering Science, Graduate school of Engineering Science,
Osaka University, Toyonaka, Osaka 560-8531, Japan}
\affiliation{CREST Research Team for Photonic Quantum Information, 4-1-8 Honmachi, Kawaguchi,
Saitama 331-0012, Japan}

\author{Nobuyuki Imoto}
%\email{}
\affiliation{Department of Materials Engineering Science, Graduate school of Engineering Science,
Osaka University, Toyonaka, Osaka 560-8531, Japan}
\affiliation{CREST Research Team for Photonic Quantum Information, 4-1-8 Honmachi, Kawaguchi,
Saitama 331-0012, Japan}

\
\date{\today}

%%%%%%%%%%%%%%%%%%%%%%%%%%%% abstract %%%%%%%%%%%%%%%%%%%%%%%%%%%%%
\begin{abstract}
We propose a realistic protocol to generate entanglement between quantum memories at neighboring nodes in hybrid quantum repeaters.
Generated entanglement includes only one type of error, which enables efficient entanglement distillation.
In contrast to the known protocols with such a property, our protocol with ideal detectors achieves the theoretical limit of the success probability and the fidelity to a Bell state, promising higher efficiencies in the repeaters.
We also show that the advantage of our protocol remains even with realistic threshold detectors. 
\end{abstract}
\maketitle

%%%%%%%%%%%%%%%%%%%%%%%%%%%%%%% intro %%%%%%%%%%%%%%%%%%%%%%%%%%%%%
If two distant parties hold quantum memories in a maximally entangled state (Bell state), 
they can freely accomplish applications such as quantum teleportation \cite{B93} and quantum key distribution \cite{E91}. 
In order to prepare their memories in a Bell state, 
optical pulses are used as media to exchange quantum information of the memories.
However, the real transmission channel for optical pulses suffers from
loss that increases with the length $l$ of the channel, which
renders direct distribution of Bell states over long distances
(e.g. $\sim 1000$km) practically impossible.
Instead, in quantum repeaters \cite{B98,D99}, a number of nodes at moderate intervals (e.g. $\sim 10$km) are set between the two parties, and many non-maximally entangled states are shared between neighboring nodes through the lossy channel.
From these non-maximally entangled states, neighboring nodes prepare a Bell state by entanglement distillation \cite{B96,D96},
and then the Bell states connecting nodes are further converted into a Bell state between the two end parties by entanglement swapping \cite{B93,Z93}.
In this way, quantum repeaters enable the distribution of Bell states over long distances through a series of `entanglement generation,' `entanglement distillation' and `entanglement swapping.'
As the first stage in quantum repeaters, entanglement
generation between neighboring nodes plays an important role in
improving the efficiency of the whole process. Thus, it is crucial to implement good
entanglement generation protocols with efficient production of high
quality entanglement to be fed to the entanglement distillation stage.

In general, the feasibility and the efficiency of entanglement generation protocols depend on the available systems for constructing quantum memories. 
The first realistic entanglement generation protocol proposed by Duan {\em et al.} \cite{D01} and the subsequent protocols \cite{B07,Z07,J07} are based on atomic-ensemble quantum memories giving off a single photon depending on the state of the atoms.
Although the protocols generate high quality entanglement, they suffer from low success probabilities.
This is due to the necessity to suppress the generation of multiple photons.
On the other hand, quantum memories used in hybrid quantum repeaters \cite{L06,S06,L08,M08} do not have such a restriction, and the repeaters are expected to be more efficient.
For example, the memory $M$ used in the first proposal of Loock {\em et al.} \cite{L06} interacts with the optical pulse $c$ in coherent state $\ket{\alpha}_c=e^{-|\alpha|^2/2} e^{\alpha \hat{a}^\dag} \ket{0}_c$
with any amplitude $\alpha$ according to 
$
\hat{U}_{\theta} (\ket{0}_M \ket{\alpha}_c)=\ket{0}_M \ket{\alpha e^{i \theta/2}}_c,\; \hat{U}_{\theta}(\ket{1}_M \ket{\alpha}_c)=\ket{1}_M \ket{\alpha e^{-i  \theta/2}}_c$, 
where the parameter $\theta$ depends on the strength of the interaction (e.g. $\theta \sim 0.01$ \cite{L06}).
Such quantum memories can be realized by individual $\Lambda$-type atoms, single electrons trapped in quantum
dots, and nitrogen-vacancy (NV) centers
in a diamond with a nuclear spin degree of freedom \cite{L06,S06}.
Until now, many authors have tried to improve entanglement generation protocols in order to achieve higher efficiencies in the repeaters \cite{L06,S06,L08,M08}.

In this paper, we provide a definitive step in the search of good entanglement generation methods for hybrid quantum repeaters by proposing a protocol with an optimal performance.
Our protocol uses linear optical elements and photon detectors, and 
if photon number-resolving detectors are available, the protocol promises to achieve the theoretical limit of performance under the requirement that the generated entanglement suffers from only one type of error. This required property makes subsequent entanglement distillation efficient \cite{M08}.
All known protocols \cite{L06,S06,L08,M08,C06,C05}, including a protocol generating entanglement with two types of errors \cite{L06,S06}, do not reach the bound, and thus our protocol is the first one achieving the bound.
In addition, even if realistic detectors are used,
our protocol shows higher performance than known realistic protocols.
Hence, it is a promising protocol to achieve more efficient hybrid quantum repeaters.

%%%%%%%%%%%%%%%%%%%%%%%%%%%%%%% main part %%%%%%%%%%%%%%%%%%%%%%%%%%%%%

%%%%%%%%%%%%%%%%%%%%%%%%%%%%%%% scheme %%%%%%%%%%%%%%%%%%%%%%%%%%%%%

Our entanglement generation protocol is illustrated in Fig.~1(a).
In what follows, we call the sender and the receiver at neighboring
nodes as Alice and Bob, respectively, who are connected via 
an optical fiber
with transmittance $T=e^{-l/ l_0 }$, where 
$l$ is the distance between the nodes.
Alice first prepares a probe pulse in a coherent state $\ket{\alpha}_{
a}$ with $\alpha\ge 0$
and a quantum memory $A$ in state 
$(e^{-i(\xi+ \zeta) }\ket{0}_{A}+e^{i (\xi+ \zeta)} \ket{1}_{A})/\sqrt{ 2}$ 
with $\zeta:=(1/2)T\alpha^2\sin{\theta}$ and $\xi:=(1/2) (1-T)\alpha^2\sin{\theta}$, where phase factors $\xi$ and $\zeta$ are chosen to offset irrelevant phases appearing later.
Alice then makes the probe pulse interact with her memory by
$\hat{U}_{\theta}$, and 
sends the output probe pulse to Bob through the fiber,
together with the local oscillator (LO).
Optical loss in the fiber is effectively described by
$
\hat{N}\ket{\alpha}_a =  \ket{\sqrt{T} \alpha}_{b_1} \ket{\sqrt{1-T} \alpha}_E,
$
where $\hat{N}$ is an isometry from input mode $a$ into output mode $b_1$ and the environment $E$.
Then, the state of Alice's memory $A$, the received probe pulse in mode $b_1$, and the environment $E$ is described by
$
\ket{\psi}_{Ab_1E}=(\ket{0}_{A}\ket{u_0}_{b_1}\ket{v_0}_{ E}
+\ket{1}_{ A}\ket{u_1}_{ b_1}\ket{v_1}_{ E})/\sqrt{2}$ with $\ket{u_j}_{b_1}:= e^{-i(-1)^j \zeta} \ket{\sqrt{T}\alpha e^{i (-1)^j \theta/2 }}_{ b_1}$ and $\ket{v_j}_E:= e^{ -i(-1)^j  \xi} \ket{\sqrt{1-T}\alpha e^{i (-1)^j \theta/2 }}_{ E}$.

The above recipe for Alice is also shared by the protocols in Refs. \cite{L06,S06,L08}, while that for Bob is not. 
In these protocols, Bob first interacts the probe pulse with his memory, and then he either performs homodyne measurement on the probe pulse (protocol I)
\cite{L06,S06} or displaces the probe pulse and conducts photon counting 
(protocol II) \cite{L08}. 
As seen below, our
protocol differs from these in the sense that 
it uses two probe pulses.
Note that the two-probe approach was also taken in other repeater protocols \cite{D01,B07,Z07,J07,C05,C06}.

In our protocol, upon receiving the probe pulse and the LO pulse, Bob first generates a second probe pulse in state 
$\ket{\sqrt{T} \alpha}_{ b_2}$ from the LO with a beamsplitter (BS2), 
and then makes it interact with his memory initialized in state 
$(e^{-i\zeta}\ket{0}_{B}+e^{i\zeta}\ket{1}_{B})/\sqrt{2}$.
Then, his memory and the second probe pulse are in state 
$\ket{\phi}_{Bb_2}=(\ket{0}_{ B}\ket{u_0}_{b_2}
+\ket{1}_{ B}\ket{u_1}_{ b_2})/\sqrt{2}$.
Bob further applies a 50/50 BS (BS3) described by $\ket{\alpha_1}_{b_1} \ket{\alpha_2}_{b_2} \to \ket{(\alpha_2 - \alpha_1)/\sqrt{2}}_{b_3} \ket{(\alpha_2+\alpha_1)/\sqrt{2}}_{b_4}$ to the pulses in modes $b_1$ and $b_2$, which is followed by a phase-space displacement $\hat{D}(-\sqrt{2T} \alpha \cos(\theta/2))$ \cite{note-d} to the pulse in mode $b_4$.
These operations correspond to the following isometry: 
$
\ket{u_0}_{ b_1} \ket{u_0}_{b_2}
\to
\ket{0}_{b_3}\ket{\beta}_{b_5},\;
\ket{u_0}_{ b_1}\ket{u_1}_{b_2} \to
\ket{-\beta}_{b_3}\ket{0}_{b_5},\;
\ket{u_1}_{ b_1} \ket{u_0}_{b_2}
\to
\ket{\beta}_{b_3}\ket{0}_{b_5},\;
\ket{u_1}_{ b_1} \ket{u_1}_{b_2}
\to \ket{0}_{b_3}\ket{-\beta}_{b_5},
$
where $\beta:=i \sqrt{2T}\alpha \sin{(\theta/2)}$.
Then, the state of the total system is described by 
$
\ket{\chi}_{ABb_3b_5E}
= \ket{0}_{b_3} (\ket{00}_{AB} \ket{\beta}_{b_5} \ket{v_0}_E+
\ket{11}_{AB} \ket{-\beta}_{b_5} \ket{v_1}_E)/2 
+\ket{0}_{b_5} ( \ket{01}_{AB} \ket{-\beta}_{b_3} \ket{v_0}_E
+\ket{10}_{AB}\ket{\beta}_{b_3}  \ket{v_1}_E )/2.
$
The pulses in $b_3$ and $b_5$ go to photon detectors D1 and D2, respectively, and Bob announces the success of the protocol when either photon detector D1 or D2, but not both, reports the arrival of nonzero photons.

%%%%%%%%%%%%%%%%%%%%%%%%%%%%%%%%%%(}1)%%%%%%%%%%%%%%%%%%%%%%%%%%%%%%%%
% }'Ì'}"ü
\begin{figure}[t ]
  \begin{center}
    \includegraphics[keepaspectratio=true,height=60mm]{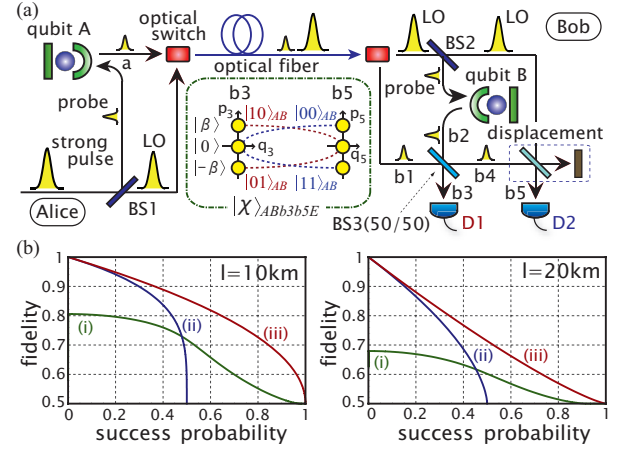}
  \end{center}
   \caption{
(a) Schematic diagram of our protocol.
(b) The performance of protocols with ideal detectors: fidelity of the obtained entanglement to a Bell state 
as a function of the success probability when $l_0=25$ km (corresponding to $\sim 0.17$ dB/km attenuation) and $\theta=0.01$, for (i) protocol I \cite{L06,S06}, (ii) protocol II \cite{L08}, and
(iii) our protocol.}
 \label{lo1}
\end{figure}
%%%%%%%%%%%%%%%%%%%%%%%%%%%%%%%%%%(}1)%%%%%%%%%%%%%%%%%%%%%%%%%%%%%%%%

Let us consider the case where D1 and D2 are ideal photon number-resolving detectors.
Since the detectors have no dark counts, the output state $\ket{\chi}_{ABb_3b_5E}$ never provokes an event where both detectors receive photons. 
Hence our protocol fails only when
the pulses in modes $b_3$ and $b_5$ are in the vacuum state
$\ket{0}_{b_3}\ket{0}_{b_5}$, which 
leads to
the success probability of $P_s(\alpha)=1-||{}_{b_3}\bra{0}{}_{b_5}\bra{0} \ket{\chi}_{ABb_3b_5E}||^2=1-e^{-2 T \alpha^2 \sin^2 (\theta/2)}.
$

The type of the generated entanglement in qubits $AB$ depends on which
detector informs how many photons have arrived. If
detector D1 announces that the number of arriving photons is odd (even but nonzero), the generated entangled state has fidelity $F(\alpha)=(1+e^{- 2 (1-T) \alpha^2 \sin^2 (\theta/2) } )/2$ to the nearest Bell state $\ket{\Psi^-}_{AB}:=(\ket{10}_{AB}-\ket{01}_{AB})/\sqrt{2}$ ($\ket{\Psi^+}_{AB}:=(\ket{10}_{AB}+\ket{01}_{AB})/\sqrt{2}$), and it is diagonalized by Bell states $\{\ket{\Psi^\pm}_{AB}\}$. Similarly, 
detector D2 informs whether the nearest Bell state to the obtained entanglement is $\ket{\Phi^{-}}_{AB}:=(\ket{00}_{AB}-\ket{11}_{AB})/\sqrt{2}$ or $\ket{\Phi^{+}}_{AB}:=(\ket{00}_{AB}+\ket{11}_{AB})/\sqrt{2}$.
These facts can be confirmed 
by simple calculations, e.g., ${}_{b_3} \bra{n} \ket{\chi}_{AB b_3 b_5 E} = \ket{0}_{b_5} (\braket{n}{-\beta} \ket{01}_{AB} \ket{v_0}_E + \braket{n}{\beta} \ket{10}_{AB} \ket{v_1}_E)/2$ for the number state $\ket{n}_{b_3}$ ($n>0$),
$\braket{n}{\beta}=(-1)^n \braket{n}{-\beta} $, and $\braket{v_1}{v_0}=e^{- 2 (1-T) \alpha^2 \sin^2 (\theta/2) } $.
Then, using a local unitary operation depending on the outcome of the detectors, Alice and Bob can transform the generated entangled state into the standard state, $F(\alpha) \ket{\Phi^{+}} \bra{\Phi^{+}}_{AB} + (1-F(\alpha)) \ket{\Phi^-} \bra{\Phi^{-}}_{AB}$.
Since the standard state includes only one type of error, they can use efficient entanglement distillation at a later stage \cite{M08}.
This property is also shared by protocol II \cite{L08} and by another protocol \cite{M08}.

In order to evaluate the potential of our protocol, we compare its
performance with protocols
I and II in Fig.~1(b), assuming ideal photon number-resolving detectors and ideal homodyne detectors.
The figure suggests that our protocol has the best performance among the protocols.
In addition, the figure shows that, in the vicinity of zero success
probability, protocol II and our protocol achieve 
a fidelity close to unity,
while protocol I does not unless $T=1$ ($l=0$).
This difference comes from the choice of different types of detectors, and it is further amplified with the increase of distance $l$:
In fact, for $l \ge 40$ km, protocol I can generate almost separable states at best \cite{L08}, but our protocol and protocol II can generate acceptable entanglement.
The better performance of our protocol 
was also supported by numerical simulations for various values of $T$.

%%%%%%%%%%%%%%%%%%%%%%%%%%%%%%% optimality %%%%%%%%%%%%%%%%%%%%%%%%%%%%%

%%%%%%%%%%%%%%%%%%%%%%%%%%%%%%% optimality %%%%%%%%%%%%%%%%%%%%%%%%%%%%%

Actually, such a high potential of this protocol is not accidental,
because it can be shown to have the maximal performance among a wide range of protocols, which generate entangled states with only one type of error. 
That is to say,
our protocol achieves the optimality of entanglement generation in qubits $AB$ among all the protocols that satisfy the following conditions:
(i) Alice prepares qubit $A$ and pulse $a$ in a state $
(\sum_{j=0,1} e^{i \varphi_j} \ket{j}_A \ket{\alpha_j}_a)/\sqrt{2}$ with 
$\{\ket{\alpha_j}_a \}_{j=0,1}$ being arbitrary coherent
states, 
and sends the pulse $a$ to Bob;
(ii) 
Upon receiving the pulse (in mode $b_1$), 
Bob may perform arbitrary operations and measurements on $b_1$, the LO,
and his memory qubit $B$, but whenever he declares success, Alice and Bob
can apply a local unitary operation $\hat{U}_A \otimes \hat{U}_B$ such that the final state of $AB$ is represented only by $\{\ket{\Phi^{\pm}}\}$
(contained in the subspace spanned by $\{\ket{\Phi^{\pm}}\}$).
Condition (i) is satisfied by protocol I, II and the others \cite{L06,S06,L08, M08,C05,C06}. 
Condition (ii) is favorable since it allows the use of efficient entanglement distillation \cite{M08}.

From condition (i), we see that the state of the system $A b_1 E$
when the pulse arrives at Bob is written by
\begin{equation}
 \ket{\psi}_{A b_1 E} = \sum_{j=0,1} \ket{j}_A \ket{u_j}_{b_1}
  \ket{v_j}_E 
 /\sqrt{2}
\label{eq:abe}
\end{equation}
with 
\begin{equation}
 (1-T) \ln|\braket{u_1}{u_0}| = T \ln|\braket{v_1}{v_0}|,
 \label{tra}
\end{equation}
where $T$ is the transmittance of the fiber.
Since the cases with $|\braket{v_1}{v_0}|=1$ are trivial, 
we assume $|\braket{v_1}{v_0}|<1$
in what follows, and we use condition (ii) and Eq.~(\ref{eq:abe}) 
to derive bounds on the success probability $P_s$ and the fidelity $F$
in terms of $|\braket{u_1}{u_0}|$ and $|\braket{v_1}{v_0}|$.
Then we use Eq.~(\ref{tra}) to determine the achievable region 
of $(P_s,F)$ for given $T$.

Let us define a phase flip channel $\Lambda_A$ on qubit $A$ by
$\Lambda_A(\hat{\rho}):=q \hat{\rho} + (1-q) \hat{\sigma}_z^A \hat{\rho}
\hat{\sigma}_z^A$ with $q:=(1+|\braket{v_1}{v_0}|)/2$ and
$\hat{\sigma}_z^A := \ket{0} \bra{0}_{A} -\ket{1} \bra{1}_{A}$.
From Eq.~(\ref{eq:abe}), we have ${\rm Tr}_{E} [ \ket{\psi}\bra{\psi}_{A b_1 E} ] = \Lambda_A(\ket{\psi_{\rm id}}\bra{\psi_{\rm id}}_{A b_1})$,
where 
$\ket{\psi_{\rm id}}_{A b_1}:= \sum_{j=0,1} e^{i (-1)^j  \varphi } \ket{j}_A \ket{u_j}_{b_1}/\sqrt{2}$ with $2 \varphi:=\arg [\braket{v_1}{v_0}]$. 
The effect of the lossy channel is thus equivalently described as
preparation of $\ket{\psi_{\rm id}}_{A b_1}$ followed by
$\Lambda_A$. Since any operation of Bob commutes with $\Lambda_A$, 
the protocol is equivalent to the following sequence:
(a) System $A b_1$ is prepared in $\ket{\psi_{\rm id}}_{A b_1}$;
(b) Bob does his operations and measurements, and leaves system $AB$
in a state $\hat{\rho}_{AB}$;
(c) $\Lambda_A$ is applied on qubit $A$.
Now condition (ii) requires that, whenever Bob declares success, there exists a unitary 
$\hat{U}_A \otimes \hat{U}_B$ 
such that 
$ \bra{ \Psi'^{\pm} }\Lambda_A (\hat{\rho}_{AB}) \ket{ \Psi'^{\pm} } =0$
with $\ket{ \Psi'^{\pm} }_{AB} := \hat{U}_A^\dag \otimes \hat{U}_B^\dag \ket{ \Psi^{\pm} }_{AB}$.
Since $\hat{\rho}_{AB}$ is positive and $0<q<1$, 
we have
$\sqrt{\hat{\rho}_{AB}} \ket{ \Psi'^{\pm} }=0$
and $\sqrt{\hat{\rho}_{AB}} \hat{\sigma}_z^A \ket{ \Psi'^{\pm} }=0$
for both $\pm$. 
Adding and subtracting these equations, we obtain 
\begin{equation}
\sqrt{ \hat{\rho}_{AB}}  \ket{x_j}_A \ket{y_{j\oplus 1}}_B = \sqrt{ \hat{\rho}_{AB}} \hat{\sigma}_z^A \ket{x_j}_A \ket{y_{j \oplus 1}}_B =0 \label{zero}
\end{equation}
for $j=0,1$, where $\ket{x_j}_A:=\hat{U}_A^\dag \ket{j}_A$ and $\ket{y_j}_B:=\hat{U}_B^\dag \ket{j}_B$.
Since $\hat{\rho}_{AB}\neq 0$, 
the set $\{ \ket{x_j}_A \ket{y_{j \oplus 1}}_B ,\hat{\sigma}_z^A
\ket{x_j}_A \ket{y_{j \oplus 1}}_B\}_{j=0,1}$ must be linearly dependent,
which only happens when
$\{\ket{x_j}_A \}_{j=0,1}$ is an eigenbasis of
$\hat{\sigma}_z^A$.

Without loss of generality, the fidelity $F$ of the final state is given by
 $F=\bra{ \Phi'^{+} }\Lambda_A (\hat{\rho}_{AB}) \ket{ \Phi'^{+} }$,
where $\ket{ \Phi'^{\pm} }_{AB} := \hat{U}_A^\dag \otimes \hat{U}_B^\dag
 \ket{ \Phi^{\pm} }_{AB}$
$=(\ket{x_0}_A \ket{y_{0}}_B\pm \ket{x_1}_A \ket{y_{1}}_B)/\sqrt{2}$.
Since $\{\ket{x_j}_A \}_{j=0,1}$ is an eigenbasis of
$\hat{\sigma}_z^A$, we have 
$\hat{\sigma}_z^A\ket{ \Phi'^{+} }=\pm \ket{ \Phi'^{-} }$.
Hence $F=q\bra{ \Phi'^{+} }\hat{\rho}_{AB} \ket{ \Phi'^{+} }
+(1-q)\bra{ \Phi'^{-} }\hat{\rho}_{AB} \ket{ \Phi'^{-}}$,
leading to 
\begin{equation}
 F\le (1+|\braket{v_1}{v_0}|)/2.
\label{eq:fbound}
\end{equation}

In order to find a bound on $P_s$,
imagine a situation where, after the steps (a)--(c) above,
Alice and Bob proceeds as follows: (d) Bob measures qubit $B$ on 
basis $\{\ket{y_k}_B\}_{k=0,1}$; (e) Alice 
measures qubit $A$ on 
basis $\{\ket{j}_A\}_{j=0,1}$. Whenever Bob has declared success,
we see from Eq.~(\ref{zero}) that the state of qubit $A$ after step (d)
should be $\ket{x_k}_A$, which is an eigenvector of $\hat{\sigma}_z^A$.
Hence Bob can certainly predict Alice's outcome $j$ in step (e).
Now if we look at the whole sequence (a)--(e), we notice that
Alice's measurement (e) can be equivalently done just after (a),
and (c) becomes redundant. Then, 
when Alice finishes steps (a) and (e),
Bob is provided with $\{ \ket{u_j}_{b_1} \}_{j=0,1}$ 
with equal {\em a priori} probabilities, from which he 
proceeds with steps (b) and (d).
At this point, he can determine the value of $j$ precisely
whenever he declares success.
Thus, the total success probability $P_s$
is no larger than that of the unambiguous state discrimination (USD),
which is known \cite{D88,I87, P88} to be $1-|\braket{u_1}{u_0}|$. Hence
we have
\begin{equation}
 P_s\le 1-|\braket{u_1}{u_0}|.
\label{eq:pbound}
\end{equation}

Combining Eqs.~(\ref{tra}), (\ref{eq:fbound}), and (\ref{eq:pbound}),
we conclude that, for given $T<1$, the performance $(P_s, F)$ of any
protocol satisfying conditions (i) and (ii) must lie within the boundary 
$\{( 1-s,(1+s^{(1-T)/T})/2)\; |\; 0 \le s \le 1\}$.
Conversely, this boundary is always achievable by our protocol with the
choice of amplitude $\alpha$ satisfying $s=e^{-2 \alpha^2 \sin^2
(\theta/2)}$.

%%%%%%%%%%%%%%%%%%%%%%%%%%%%%%%%%%(}2)%%%%%%%%%%%%%%%%%%%%%%%%%%%%%%%%
% }'Ì'}"ü
\begin{figure}[tb]
  \begin{center}
    \includegraphics[keepaspectratio=true,height=28mm]{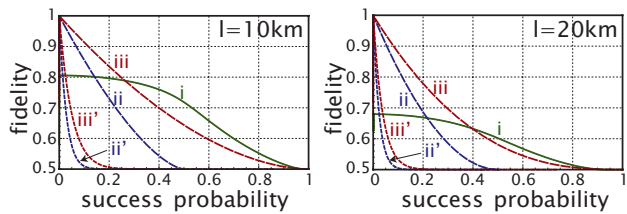}
  \end{center}
 \caption{ The performance of protocols with realistic detectors:
(i) protocol I with an ideal homodyne detector, 
(ii) protocol II with a TD1 $(\eta=0.89,\;\nu=1.4 \times 10^{-6})$, (ii') protocol II with a TD2 $(\eta=0.12,\;\nu=3.2 \times 10^{-7})$, 
(iii) our protocol with TD1s, (iii') our protocol with TD2s.
}
 \label{lo1}
\end{figure}

%%%%%%%%%%%%%%%%%%%%%%%%%%%%%%%%%%(}2)%%%%%%%%%%%%%%%%%%%%%%%%%%%%%%%%

Finally, we show that our protocol shows high performance even if we replace the photon number-resolving detectors with threshold detectors (TDs) that just report the arrival of photons and do
not tell how many of them have arrived.
We represent quantum efficiency and mean dark count of the detector as $\eta$ and $\nu$, respectively.
The function of a TD is \cite{SMB98} represented by the following POVM elements:
$\hat{E}_{\rm nc}=\sum_{m=0}^{\infty}e^{-\nu}(1-\eta)^{m}\ket{m}\bra{m},\; \hat{E}_{\rm c}=\hat{I}-\hat{E}_{\rm nc}$,
where $\hat{E}_{\rm c}$ ($\hat{E}_{\rm nc}$) corresponds to an event reporting the arrival (non-arrival) of photons.
When the used TDs are ideal ($\eta = 1$, $\nu=0$), 
the generated state has only one type of error and has fidelity $(1+e^{- 2 \alpha^2 \sin^2(\theta/2)} )/2$ to the nearest Bell state.
The success probability is the same as that with ideal photon number-resolving detectors. 
For the realistic values of $(\eta,\nu)$, 
we numerically calculated the performance $(P_s,F)$ of our protocol,
which is shown in Fig.~2.
Note that the chosen values $(\eta, \nu)$ are typical for currently available detectors, e.g., TES (superconducting transition-edge sensors) \cite{D08} and APD (avalanche photodiode) \cite{G04}.
The dark counts of such detectors increase the types of errors occurring in generated entanglement.
However, such additional errors occur with a small probability $\sim \nu (P_s^{-1} -1)+{\cal O} (\nu^2)$, and hence can be neglected.
To evaluate the performance of our protocol, we also plotted the performance of protocol I with an ideal homodyne detector, and that of protocol II with its photon number-resolving detector replaced by TD1 and TD2.
The figure shows that our protocol has higher efficiency than protocol II. We see that there is a region where the performance of protocol I exceeds that of ours, but this region decreases with the increase of distance $l$. 
Hence, we can safely say that our protocol outperforms the other protocols in the cases where long-distance and/or high quality entanglement generation is required.
It is also worth to mention that entanglement generated by protocol I always includes two types of non-negligible errors, which will affect its performance in the entanglement distillation stage.

In conclusion, we have proposed a realistic entanglement generation
scheme for hybrid quantum repeaters, which outperforms
the generation schemes proposed so far.
More importantly, we have shown that our protocol achieves
the optimal performance among all the schemes satisfying
a couple of plausible conditions [(i) and (ii) above].
Due to these conditions, our argument does not exclude the
possibility of a better protocol starting with an asymmetric
state of the sender's quantum memory, or one resulting in
 multiple types of errors possibly combined with a novel
realistic distillation protocol that has yet to be discovered.
Although such a protocol, if any, may be quite interesting,
we believe it is unlikely and our protocol is indeed the
best scheme for hybrid quantum repeaters.
The performance of our scheme will also serve as a benchmark when
one compares other types of quantum repeaters to
hybrid quantum repeaters.

We would like to thank Kae Nemoto for communicating a detail of the paper \cite{M08}.
This work was supported by JSPS Grant-in-Aid for Scientific
Research (C) 20540389 and by MEXT Grant-in-Aid for the Global COE Program and
Young scientists (B) 20740232.
K.A. and R.N. are supported by JSPS Research Fellowships for Young
Scientists.

%%%%%%%%%%%%%%%%%%%%%%%%%%%%%%% reference %%%%%%%%%%%%%%%%%%%%%%%%%%%%%


\begin{thebibliography}{99}
\bibitem{note-d}
This displacement is unnecessary
if the controlled displacement, $\hat{V}_{Mc}(\ket{j}_M\ket{0}_c)=\ket{j}_M\ket{(-1)^j\alpha'}_c\;(j=0,1)$, between the memory $M$ and the optical mode $c$ is available to Alice and to Bob \cite{M08}.
\end{thebibliography}
\end{document}